# Minimum energy and photon content in PT symmetric metamaterials

J. B. Pendry[1*] and S. A. R. Horsley[2]

[1]The Blackett Laboratory, Department of Physics, Imperial College London, London SW7 2BW, United Kingdom.
[2]School of Physics and Astronomy, University of Exeter, Stocker Road, Exeter, EX4 4QL, UK.

**ABSTRACT**. In the context of waves in space–time modulated materials, we ask two questions 'how much energy does it cost to break time reversal symmetry and transition to a PT symmetric state?' and 'can a PT symmetric system have a 'ground state' in the sense that no photons are present?'. Our model system is a periodic metamaterial set in virtual motion with velocity $c_g$ to become a space-time crystal. We find that the expectation of energy content is always increased on breaking symmetry. At the same time breaking T-symmetry introduces photon-pairs even when we start from a T-symmetric ground state empty of photons, except in certain pathological examples which we describe. For a range of velocities, PT symmetry is broken so that energy must be continuously invested to preserve motion, creating a trail of photon pairs.. Here energy must be continuously invested to preserve motion. We make an analogy with acoustic radiation generated from breaking the sound barrier.

## I. INTRODUCTION

PT symmetric systems are described by a Hamiltonian that is not Hermitian but nevertheless has real eigenvalues [1-3]. These do not correspond to energy but to some other conserved quantity the nature of which depends on details of the system. In a quantum system [4-6] there is an expectation value for the energy, $\langle U \rangle$, but actual measurements will vary from one identically prepared system to another. The question arises, but to our knowledge has not so far been answered, of what is the minimum value for $\langle U \rangle$? In the case of an optical system we can also ask whether states can be prepared having zero photon content, the PT symmetric equivalent of a ground state, $\langle N \rangle = 0$, which of course implies that $N = 0$ is an eigenstate.

We explore these questions in the context of optical materials structured in space and time [7]. We consider the case of a structure set in uniform motion with a velocity $c_g$: the permittivity and permeability have the form $\varepsilon(x - c_g t), \mu(x - c_g t)$. By this we do not mean that the material itself moves, rather that $\varepsilon, \mu$ are locally modulated to simulate virtual motion: by analogy with a wave on the sea that appears to move but the water does not.

Such systems can have a Hamiltonian that is PT symmetric with eigenvalues that are real but are not to be interpreted as energy content. In the system we shall examine, eigenvalues represent the total spin of all photons in the system and therefore spin is conserved [8].

We confirm that at lower velocities the system is PT symmetric but there is a minimum energy required to start motion, increasing with $c_g$. Once this energy is invested the system continues without further input of energy. In contrast there is a regime of broken PT symmetry associated with systems where at some point the local velocity of light is the same as that of the grating. Here we find constant emission of photon pairs with opposite spin (conserving total spin) and a constant supply of energy is needed to keep the system in motion: a sort of quantum friction [9,10]. As $c_g$ approaches the symmetry-breaking transition, energy needed for motion diverges in our model, though including dispersion would give a finite but large number. An airplane breaking the sound barrier would be a good analogy, the 'sonic boom' comprising a shower of photon pairs.

In the PT symmetric region photon content is defined only as an expectation value which in general is non-zero as photon pairs fluctuate in and out of existence. However there are special cases where photon number commutes with the Hamiltonian and is therefore conserved and zero photon eigenstates can be defined.

## II. PT–SYMMETRY IN A MOVING GRATING

Our model is an elementary case of a metamaterial in uniform motion,

$$nc_0^{-1} = \varepsilon = \mu = c_1^{-1}/(1 + 2\alpha \cos(gx - \Omega t)) \quad (1)$$

*Contact author: j.pendry@imperial.ac.uk

where $n$ is the refractive index, $c_0$ is the speed of light in vacuum. The velocity is given by $c_g = \Omega/g$ and the impedance $\sqrt{\mu/\varepsilon}$ is constant eliminating back scattering and simplifying calculations. This profile is invariant under a simultaneous inversion in space $x \to -x$ and time reversal $t \to -t$. The model is a much used template for study of virtual motion in the context of classical waves [7,11-15]. We shall mention its limitations in our conclusions.

In the absence of back scattering, time evolution factorizes into forward and backward waves. As discussed in Ref. [6] only waves travelling in the same direction as $c_g$ are of interest as these interact most strongly with the moving structure. Making this assumption, the displacement field evolves in time according to [16],

$$\partial_x n^{-1} D = -\partial_t D \qquad (2)$$

Since $n$ is spatially periodic in a co-moving Galilean frame $X = x - c_g t$, the eigenstates of (2) are defined by a Bloch wavevector k, and frequency ω. In this frame (2) becomes a non–Hermitian eigenvalue problem for $\omega(k)$

$$-i\partial_X \left(n^{-1} - c_g\right) D = \omega D \qquad (3)$$

which is symmetric under simultaneous inversion of the $X$ axis and complex conjugation (we assume $n$ is an even function, as in Eq. (1)). If $\left(n^{-1} - c_g\right)$ has the same sign throughout space, (3) can be transformed to a Hermitian problem in terms of a new field $D' = \left(n^{-1} - c_g\right)^{1/2} D$, showing that, in the PT–symmetric phase, the eigenvalue $\omega$ is real.

We can show this general behavior explicitly for the profile (1). Using the solution to (2) (given in [16] and reproduced in Eq. (A.6) of the appendix) and demanding the wave accumulate a phase $\exp(2\pi i k/g)$ over a single unit cell, the dispersion relation is found to be

$$\omega_m = c(k + mg)$$
$$= \left(c_g + \sqrt{(c_1 - c_g)^2 - (2c_1\alpha)^2}\right)(k + mg) \qquad (4)$$

Where $m$ is an integer,

We express $\omega$ in the observer rather than Galilean frame. Note the branch points at $c_g = c_1(1 \pm 2\alpha)$.

*Contact author: j.pendry@imperial.ac.uk

These branch points arise from the existence of locations in the moving index profile (1) where the local wave velocity equals $c_g$, exactly the condition that prevents us from transforming (3) to a Hermitian problem. In these cases there is an unstable accumulation of energy over time [6, 16, 17]. For a grating velocity $c_g$ between these values, the evolution of the $D$ field breaks the PT symmetry of the system.

In the observer's frame many frequencies are found differing by multiples of $\Omega$. Nevertheless, the eigenstates have the simple form (again see [16] and Eq. (A.6) of the appendix) $\exp(ikf) \times \partial_X f(X)$ where,

$$|D| = \partial_X f(X) = \left(c_0 n^{-1} - c_g\right)^{-1}, \qquad (5)$$

showing that lines of $|D|$ are most compressed where the local velocity, $c_0 n^{-1}$, comes close to $c_g$ and that $|D|$ is conserved. Note that $|D|$ is independent of the wave vector. A typical eigenstate is plotted in FIG. 1.

### III. GROUND STATE ENERGY AND PHOTON CONTENT

We now calculate the properties of the quantum ground state of the electromagnetic field in this space–time varying medium. To find the lowest expectation value of the energy we start from $\alpha = 0$ in Eq. (1), initially eliminating time variation from the medium. For a state with wavevector $k$ we trivially identify the zero-photons ground state energy as $U_0 = \hbar c_1 k/2$. To access the moving state we slowly increase $\alpha$. This process conserves the wavevector. If the system is large but finite and the process is slow compared to frequency spacing between states, each state will continuously evolve into an eigenstate for the current value of $\alpha$ and minimise introduction of photons.

This approach allows us to calculate the eigenstate at each stage of the process but not to normalize it. Conservation laws [16] for the lines of force solve the problem. The integral,

$$\frac{g}{2\pi} \int_0^{2\pi/g} |D_k(\alpha = 0, X)| dX = \sqrt{\frac{n(\alpha = 0)U_0}{c_1}} \qquad (6)$$

is conserved as $\alpha$ is increased. The fields in eq. (6) have been normalized to start with the static zero-point energy content, $U_0$,

$$U_0 = \frac{c_1}{n(\alpha=0)}|D_k(\alpha=0)|^2 \quad (7)$$

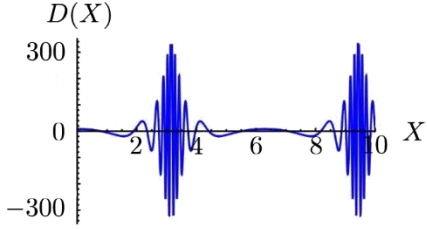

FIG. 1. Plot of $\mathrm{Re}\,D(X)$ for, $\alpha=0.025, g=1, c_1=1, c_g=0.947, k=10$. A space–time varying refractive index profile such as that given in Eq. (1) causes the electromagnetic field to concentrate around regions where the wave speed is closest to the synthetic velocity $c_g$. Note: $c_g$ is close to the singularity at $1-2\alpha=0.95$.

We may now calculate enhancement of the zero point energy by evaluating eq. (6) for non-zero $\alpha$,

$$U = \frac{g}{2\pi}\int_0^{2\pi/g} |D_{n,k}(\alpha,X)|^2 \frac{c_1}{n(\alpha,X)} dX \quad (8)$$

Using Eqs. (5-8) the energy cost of motion is calculated and plotted against $c_g$ in the top panel in FIG. 2. Here we neglect the contribution of left–going modes, which although important, does not diverge at the PT–symmetry breaking point. A more detailed calculation based on a quantization of the field is given in the appendix, leading to the same result and showing that, as demonstrated in Fig. 2a, the ground state energy $U$ of each mode diverges as we approach the symmetry broken phase where $\omega_m(k)$ in Eq. (4) becomes complex.

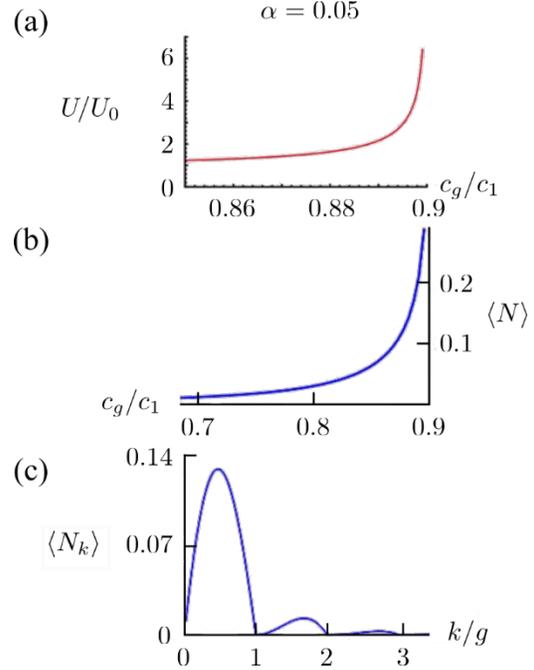

FIG. 2. (a) Plot of $U/U_0$ versus $c_g/c_1$. For these parameters ($\alpha=0.05, g=1.0$) the system enters the PT symmetry-broken regime at $c_g/c_1=0.9$. (b) Average number of photons added to the ground state versus grating velocity for $k=0.7g$. (c) Average number of photons added versus wave vector, $c_g=0.89c_1$. Whenever the wave vector is an integer multiple of $g$, no photons are created. Note that in this model which neglects dispersion, enhancement is the same for all states irrespective of frequency.

Even though there is no physical motion of the medium, the field energy (8) increases from the stationary system to the singularity at $c_g/c_1=1-2\alpha=0.9$ where PT symmetry is broken. Eq. (4) shows that $\omega$ becomes complex for $c_1(1-2\alpha)<c_g<c_1(1+2\alpha)$ indicating a continual increase in energy. In this PT–broken phase, input of energy is required to keep the system in motion, a sort of quantum friction. It was shown elsewhere [6] that photon pair production is responsible for the increase in energy.

Next we calculate the number of photons present.

*Contact author: j.pendry@imperial.ac.uk

As shown in the appendix (see also Ref. [16]), in the rest frame of the index profile, the modes of our system have a conserved frequency ω. The Hamiltonian in this frame can therefore be written as

$$\hat{H} = (4\pi)^{-1} \int d\omega\, \hbar\omega \left[ \hat{a}_\pm^\dagger(\omega)\hat{a}_\pm(\omega) + \hat{a}_\pm(\omega)\hat{a}_\pm^\dagger(\omega) \right]$$

where the "$\hat{a}_\pm$" are the annihilation operators for right/left going modes. This indicates that the number of photons is conserved in the rest frame of the medium, and - assuming the index profile becomes uniform at infinity - the vacuum state can be defined as usual as $\hat{a}_\pm |0\rangle = 0$. This all holds, provided that PT–symmetry is not broken, at which point the modes can no longer be defined over the whole of space.

Meanwhile, in the frame where the index profile is in motion, the Hamiltonian must give rise to the transformed Heisenberg equations e.g. $\partial_t \hat{A} + c_g \partial_x \hat{A} = i\hbar^{-1}[\hat{H}, \hat{A}]$. The Hamiltonian in this frame $\hat{H}'$ thus has an additional term proportional to the Minkowski momentum

$$\hat{H}' = \hat{H} + \frac{c_g}{2} \int dx \left[ \hat{D}\partial_x \hat{A} + \partial_x \hat{A}\hat{D} \right]$$
$$= \varepsilon_0^{-1} \int dx\, n^{-1} \hat{D}^2 \tag{9}$$

This additional term is not proportional to the number operator and therefore the photon number in the $|0\rangle$ state is non–zero in the $x,t$ coordinate system. This is an example of the known reference frame dependence of particle number found in e.g. the Fulling–Davies–Unruh effect [18, 19].

To count the photons in the $|0\rangle$ state we move to Fourier space in order to identify each frequency component of the eigenstates $D_{m,k}(x,t)$ ($k$ the Bloch vector and $m$ the band index). These Fourier components are defined as

$$d_{m,k,m'} = (2\pi/g) \int_0^{2\pi/g} dx\, D_{m,k}^* e^{-i(k+m'g)x}$$

where $D_{m,k}$ is normalized in the same way as in Eq. (6). As our model system is impedance matched, the displacement and magnetic fields are related, as they are in free space, by $B = \pm(c_0\varepsilon_0)^{-1} D$. We can thus imagine instantaneously "turning off" the index profile and, without this leading to any additional excitation, measuring the photon content in terms of these Fourier components

$$\langle N_{m,k} \rangle = \langle 0|\hat{a}_{m,k}^\dagger(\omega)\hat{a}_{m,k}(\omega)|0\rangle = \frac{2}{\hbar|\omega_{m,k}|\varepsilon_0} \sum_{m'} |d_{m,k,m'}|^2 \tag{10}$$

where $\hat{a}_{m,k}$ is an annihilation operator for the free space mode with wave–vector $k + mg$ and frequency $\omega_{m,k}$. This tells us that we can always count the number of photons in our system by dividing the square of the Fourier coefficients by $\hbar|\omega + m\Omega|$, and summing over all Fourier components. This approach is described further in [20, 21]. Note that if we make the same calculation except using $\hbar(\omega + n\Omega)$, i.e. without taking the modulus, we calculate the spin. It has been shown [8] that spin is conserved in this system. It follows that any increase in photon number is associated with transitions between positive and negative frequencies during the increase of $\alpha$.

The middle panel in FIG. 2 shows the total number of photons produced as the grating velocity approaches $c_1$, the lower critical velocity. Comparing the middle and upper panels we note that photon production makes a minimal contribution to energy which is mainly vested in modifying the original ground state.

Dependence of photon production on wave vector is presented in the bottom panel of FIG. 2. Photon production is heavily biased towards low values of $k$ and is strictly zero when $k$ is an integer multiple of $g$. To understand this result we note that photons can only be created in $\pm$ frequency pairs which in the absence of back scattering is the only way to conserve spin. Matrix elements of the Hamiltonian coupling frequencies are proportional to frequency [21] and frequency changes in single steps of $\Omega$ in consequence of eq. (1). If $\omega = m\Omega$ the route to negative frequency passes through $\omega = 0$, and is therefore barred: no photons can be created. Furthermore starting from a large positive value of $\omega$ means a longer journey to $\omega = 0$ and matrix elements, weighted by frequency as they are, favor migration to even larger $\omega$.

This result opens the possibility of systems which have PT-zero point energy and no photons.

Consider a wave on a ring, in which the field obeys periodic boundary conditions requiring $k = 2\pi n/L$. Then choosing $g = 2\pi/L$ in Eq. (1), no state can have photons added.

*Contact author: j.pendry@imperial.ac.uk

With the exception of this pathological case we argue that even in the lowest energy state, photons will always be present. Suppose that instead of creating states of the moving system by slowly acting on the initial ground state, we suddenly start the motion. Immediately after the start there are still no photons present. This state is not an eigenstate of the time evolution operator but can be expanded in terms of the complete set of eigenfunctions. In reciprocal space this looks like,

$$D_{m,k} = \sum_{m'} F_{m'm} e^{i(k+mg)x - i(\omega_{m'} + m\Omega)t} \quad (11)$$

where subscript $m'$ labels the eigenfunctions and $m$ the Fourier component of that eigenvector. The eigenfunctions will be in phase only at points in time when they sit in that exact relationship to the grating. At other times they will represent a state with non-zero photon expectation value. On average the photon occupation will be the sum of the component eigenvalue contributions. All eigenstates have the same velocity, $c$, and have velocity $(c - c_g)$ relative to the grating. Therefore photon content will oscillate in time with a periodicity $2\pi/(g(c - c_g))$ where $2\pi/g$ is the grating period.

## IV. CONCLUSIONS

Our simple model demonstrates the main features of energy and photon minimization in a system that transitions from separate T and P symmetries to a combined PT symmetry. Namely that it always costs energy, even if the starting point is the ground state. During the process, photon production cannot be avoided except for certain special cases, one of which we have demonstrated. This necessary investment of energy means that the systems we consider here cannot spontaneously crystalize into the PT symmetric state.

In the velocity range $c_1(1 - 2\alpha) < c_g < c_1(1 + 2\alpha)$, PT symmetry is broken and a regime of friction ensues. In our model to cross this threshold requires an infinite investment of energy a consequence of neglecting dispersion. Nevertheless even in a more sophisticated model crossing the threshold would require major investment of energy.

The main limitation of our model is absence of frequency dispersion in $\varepsilon, \mu$ which is always present in real systems. This will remove infinities from the solutions due to infinite compression of waves. This cannot happen when dispersion is present. We choose to neglect back scattering which even outside the transluminal region can break PT symmetry by introducing band gaps in $k$.

*Contact author: j.pendry@imperial.ac.uk

## ACKNOWLEDGMENTS
J.B.P. and S.A.R.H. acknowledge support from the EPSRC via the META4D Program me Grant (EP/Y015673/1), and S.A.R.H. acknowledges additional support through the EPSRC grant "Photon management in dynamic complex scattering media" (EP/Z535928/1).

## APPENDIX

Here we provide more details to show how the electromagnetic field can be quantized in this moving index profile and the calculation of the ground state energy. In the case where $\varepsilon = \mu = n$, Maxwell's equations reduce to,

$$\frac{\partial E}{\partial x} = \frac{\partial B}{\partial t}$$
$$\frac{\partial}{\partial x}\left(n^{-1}B\right) = c_0^{-2} \frac{\partial}{\partial t}(nE) \quad (A.1)$$

where we assume propagation along the x–axis and $\mathbf{E} = E\mathbf{e}_z$ and $\mathbf{B} = B\mathbf{e}_y$. In terms of the potentials, the first of the above equations is trivial whereas the second reduces to,

$$\left(\frac{\partial}{\partial x} - \frac{1}{c_0}\frac{\partial}{\partial t}n\right)\left(\frac{1}{n}\frac{\partial}{\partial x} + \frac{1}{c_0}\frac{\partial}{\partial t}\right)A = 0 \quad (A.2)$$

Considering only right-going waves, which propagate with the motion of the index profile we can take the right-most part of this factorized wave equation,

$$\left(\frac{\partial}{\partial x} + \frac{n}{c_0}\frac{\partial}{\partial t}\right)A_+ = 0. \quad (A.3)$$

As described in the main text, in the co–moving frame $X = x - c_g t$, the refractive index is time independent and the above wave equation can be written for a fixed frequency $\omega$,

$$\frac{\partial A_+}{\partial x} = \frac{i\omega}{c_0}\frac{n}{1 - nc_g/c_0}A_+ \quad (A.4)$$

with the solutions,

$$A_+(X,t) = \exp\left[\frac{i\omega}{c_0}\int^X \frac{n(X')dX'}{1 - n(X')c_g/c_0}\right] \quad (A.5)$$

The corresponding displacement field $D_+$ is defined in terms of a time derivative of the vector potential in the observer frame. Thus, in terms of the co-moving frame coordinates,

$$D_+(X,t) = -\varepsilon_0 n \left( \frac{\partial A_+}{\partial t} + c_g \frac{\partial A_+}{\partial X} \right)$$

$$= \frac{i\omega n \varepsilon_0}{1 - nc_g/c_0} \exp\left[ \frac{i\omega}{c_0} \int^X \frac{n(X')dX'}{1 - n(X')c_g/c_0} - c_0 t \right] \quad (A.6)$$

In terms of the co-moving frame coordinates we have thus found the modes of the system, with $-D_+$ being the conjugate momentum to the vector potential $A_+$. We write these operators as,

$$\hat{A} = \hat{A}_+ + \hat{A}_-$$
$$\hat{D} = \hat{D}_+ + \hat{D}_- \quad (A.7)$$

(with '$\pm$' indicating right/left propagation). The quantum theory of light requires the following commutator between these operators,

$$\left[ \hat{A}(X,t), -\hat{D}(X',t) \right] = i\hbar \delta(X - X') \quad (A.8)$$

We now look for the expansion of $\hat{A}_+$ and $\hat{D}_+$ such that this commutation relation is fulfilled. Without assuming any periodicity we can write,

$$\hat{A}_+ = \int_0^\infty \frac{d\omega}{2\pi} \sqrt{\frac{\hbar}{2\varepsilon_0 \omega c_0}} e^{\left[\frac{i\omega}{c_0}\int^X \frac{n(X')dX'}{1-n(X')c_g/c_0} - c_0 t\right]} \hat{a}_+ + \text{h.c.} \quad (A.9)$$

the displacement field operator follows from Eq. (A.6). Assuming $\left[ \hat{a}_+(\omega), \hat{a}_+^\dagger(\omega') \right] = 2\pi \delta(\omega - \omega')$, the commutation relation between $\hat{A}_+$ and $\hat{D}_+$ is,

$$\left[ \hat{A}_+(X,t), -\hat{D}_+(X',t) \right] = \frac{i\hbar}{2} \delta(X - X') \quad (A.10)$$

in agreement with the required commutation relation. The factor of one half arises because we are only considering right-going waves. Performing an identical analysis for left–going waves produces the same result, with the sum equaling (A.8). Combining (A.9) and (A.6) shows that, as described in the main text, each mode in the expansion of the displacement field has a fixed normalization $\sqrt{\hbar/(2\omega \varepsilon_0 c_0)}$ which is independent of the grating velocity $c_g$.

The argument given in the main text uses the fact that the normalization of the modes (A.6) necessary to fulfill the commutation relation (A.8) requires a fixed amplitude independent of both the index profile and the grating velocity. This normalization leads to an energy for each mode that does depend on the grating profile and velocity. This means that there is an energetic cost for setting the grating into motion: for making the transition between a time symmetric and a parity-time symmetric vacuum state.

From the above equation for $A_+$ (A.3) we can see that, for right–going waves the magnetic field and the displacement field are related as follows,

$$B_+ = -\frac{1}{c_0 \varepsilon_0} D_+ \quad (A.11)$$

(an identical relation holds in free space). Therefore the energy (in the observer's frame, where the grating is in motion) can be written as

$$U = \frac{1}{2} \int dx \left( \varepsilon_0 \varepsilon E_+^2 + \mu_0 \mu H_+^2 \right)$$

$$= \frac{1}{2} \int dx \left[ \frac{D_+^2}{n\varepsilon_0} + \frac{B_+^2}{n\mu_0} \right] \quad (A.12)$$

$$= \frac{1}{\varepsilon_0} \int dx \frac{D_+^2}{n}$$

Taking the case where the index profile is uniform at infinity, the modes in Eq. (A.9) reduce to plane waves at large $|X|$, with the Doppler shifted dispersion relation $n(\omega + c_g k)/c_0 = k$ due to the use of a Galilean transformation. The vacuum state can thus be defined by $\hat{a}_+ |0\rangle = 0$, and the ground state energy (A.12) is,

$$\langle 0| \frac{1}{\varepsilon_0} \int dx \frac{\hat{D}_+^2}{n} |0\rangle = \int dX \int_0^\infty \frac{d\omega}{2\pi} \frac{\hbar \omega}{2 c_0} \frac{n}{\left(1 - nc_g/c_0\right)^2} \quad (A.13)$$

an energy which increases with $c_g$ and diverges when the index profile contains points where $c_g = c_0/n$.

Note that when the medium is homogeneous and its velocity is zero $c_g = 0$, the integral over frequency can be replaced with one over wave–vector $dk = nd\omega/c_0$, and (A.13) reduces to the usual result for the vacuum energy. When restricted to a single mode and single period of the moving index profile, this is expression (8) given in the main text.

*Contact author: j.pendry@imperial.ac.uk

*Contact author: j.pendry@imperial.ac.uk